# Magnetodielectric coupling in a triangular Ising lattice

Natalia Bellido, Charles Simon, Antoine Maignan

Laboratoire CRISMAT, UMR 6508 CNRS ENSICAEN, 14050 CAEN, France.

Abstract:

**Dielectric constant measurement under magnetic field is an efficient technique to study the coupling between charges and spins in insulating materials. For magnetic oxides, the geometric frustration is known to be a key ingredient to observe such a coupling. Measurements for the triangular Ising-like cobaltite $Ca_3Co_2O_6$ have been made. Single crystals of $Ca_3Co_2O_6$ are found to exhibit a magnetodielectric effect below $T_N=25K$ with a peak in the $\varepsilon(H)$ curve at the ferri to ferromagnetic transition. This relation between $\varepsilon$ and magnetization has been modelized by using two order parameters in an energy expansion derived from the Landau formalism and the fluctuation-dissipation theorem. This result emphasizes the great potential of insulating transition metal oxides for the search of magnetodielectric effect.**

**I Introduction**

The search for multiferroic oxide materials is a very emerging topic with perspectives of applications in the domain of memories. However, the coupling between electrical polarisation and magnetization appears to be experimentally complex and the coexistence of these antagonist properties limits the number of candidates [1, 2, 3, 4]. A natural way to evidence a coupling between electrical charges and magnetic moments consists in magnetodielectric measurements. Recently, such a technique has been successful to reveal small changes in local spin states under external magnetic field application as for $SeCuO_3$ [5], $Dy_2Ti_2O_7$ [6] or $Co_3V_2O_8$ [7]. For this class of materials, the lack of constrains related to non centrosymmetry as for multiferroics, offers a much broader range of possible compounds. Furthermore, the observation in magnetodielectrics of a coupling between the dielectric constant and the magnetization has allowed a modelling based on the Landau theory of phase transitions to be proposed for YMnO$_3$ [8] and other compounds [9,10]. The ingredients of this model are expressed in the free energy by considering the free energy of an antiferromagnet, the free energy of a ferroelectric, the coupling with external magnetic and electric fields and the coupling between both order parameters (magnetisation and polarisation). A common feature of the best prototypical magnetodielectrics is the existence of the geometric frustration coming from tetrahedra in pyrochlores (Dy$_2$Ti$_2$O$_7$) [6], from the staircase kagome lattice (Co$_3$V$_2$O$_8$) [7] or from triangular lattice (YMnO$_3$ [8] or CuFeO$_2$ [11]).

Among the triangular lattices, the Ising-like ferromagnetic chains coupled antiferromagnetically set on a hexagonal network of Ca$_3$Co$_2$O$_6$ offers an ideal case to test the validity of the model based on the Landau theory for two order parameters, antiferromagnetic (interchain coupling) and ferromagnetic (intrachain coupling) [12, 13, 14]. In this phase, the existence of trigonal prisms characterized by a strong magneto-crystalline anisotropy is believed to be responsible for the Ising like character of the [CoO$_6$]$_\infty$ chains. Below the ordering temperature ($T_N$=25K), upon magnetic field application along the chains axis, the magnetization exhibits first a plateau corresponding to 1/3 of the saturation magnetization ($M_S$) for which 2/3 of the chains are antiferromagnetically [15] coupled and at ~ 3.6 T, an abrupt jump is observed from 1/3 of $M_S$ to $M_S$ [16,17]. At lower temperatures, secondary steps in magnetization appear equally spaced by 1.2T. Different models have been proposed in order to explain the nature of these magnetization jumps [18,19,20,21]. In the following, we report on measurements of the magnetodielectric coupling in Ca$_3$Co$_2$O$_6$ single crystals and on the modelling of this effect.

**II Experimental**

Single crystals of Ca$_3$Co$_2$O$_6$ were grown according to the method previously described [22]. This procedure leads to crystals having a needle-like shape (with the c axis along the longest dimension). Measurements of capacitance require the homogeneous distribution of charges in two parallel surfaces of the sample. Several single crystals where relieved for their suited geometries: needle shaped objects of 3mm along the chains direction and of hexagonal section of 0.05mm$^2$. Two large parallel faces of the crystals were covered with silver paste in order to apply an ac electric field perpendicular to the chains. Magnetic field was oriented along the chains. To obtain the values for the capacitance, complex impedance measurements were performed using a commercial AG4284A LCR-meter. In order to measure in a magnetic field, a sample holder with four coaxial cables was designed. This set-up allows measurements of capacitance in a Quantum Design Physical Properties Measurement System.

Resistivity measured perpendicular to the chains direction in $Ca_3Co_2O_6$ increases when decreasing temperature from the 2kΩ·cm at room temperature, reaching non-measurable values (R>$10^{11}$Ω) at about 100K. Magnetoresistance has also been measured at high temperatures, with external magnetic field up to 14T, but no effect has been revealed. Thus, this system is an insulator below 150K and no magnetoresistance or space charges might contribute to our values of dielectric constant at low temperatures. At first, the effect of the magnitude and frequency of the ac electric field was checked and the following values of 10mV and 100 kHz where chosen in order to warrant a lack of frequency dependence. The temperature range measurements is limited to T<150K corresponding to dielectric losses smaller than $10^{-2}$.

**III Experimental results**

Dielectric constant is rather constant down to $T_N$ and then presents an abrupt decrease at the antiferromagnetic transition starting at $T_N$=25K. (fig. 1), as already reported in many antiferromagnetic compounds such as $YMnO_3$ [23] and $BiMnO_3$ [9]. In order to analyse this type of behaviour, the classical derivation of the free energy F has been used:

$$F = F_{AFM}(L) + \alpha P^2 - EP + \gamma P^2 L^2 \qquad (1)$$

where $F_{AFM}(L)$ is the free energy of an antiferromagnetic compound as function of the antiferromagnetic order parameter L (the alternate magnetization) : $F_{AFM}(L) = a(T-T_N)L^2 + bL^4$ and a,b,α,γ are the temperature independent Landau expansion coefficients. P and E are the polarization and the electric field respectively. We have included two terms which represent the induced polarization due to the applied electric field and one additional term coupling magnetic order parameter to polarization [23] was also included. Parameter γ which couples P and L is originating from the LS coupling which is the only term in the microscopic Hamiltonian which couples these quantities. This was very often introduced by the Dzyloshinskii-Moriya approximation as it was done for example in [11].

Electric susceptibility can be obtained as the inverse of the second derivative of the free energy respect to polarization:

$$\frac{1}{\chi_e} = \frac{\partial^2 F}{\partial^2 P} = \alpha + \gamma L^2 \qquad (2)$$

Consequently, the dielectric constant can be developed for small $\gamma L^2$ values:

$$\varepsilon = \chi_e + 1 = \frac{1}{\alpha + \gamma L^2} + 1 \approx 1 + \frac{1}{\alpha} - \frac{\gamma \langle L \rangle^2}{\alpha^2} \qquad (3)$$

The value of $<L^2>$ can be measured by neutron scattering (this is directly proportional to antiferromagnetic peaks such as (001) [12]). This was already reported in polycrystals [13] and singlecrystals [24]. In the inset of figure 1, we have reported the data from reference 13. The experimental agreement is more qualitative than quantitative. In particular, the decrease of $<L^2>$ observed at around 15K is not observed in $\varepsilon$.

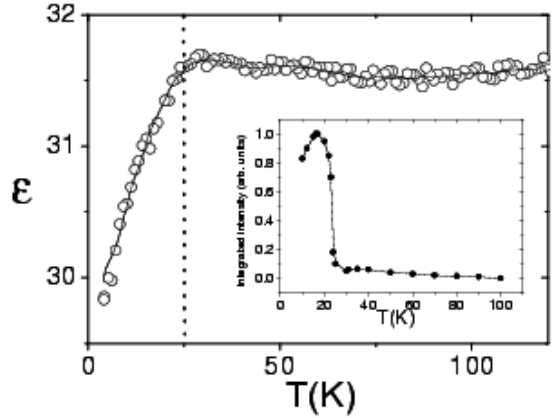

*Figure 1 : Temperature dependence of dielectric constant shows a decrease when long range magnetic order takes place ($T_N=25K$), i.e. when the antiferromagnetic order parameter (L) appears. This magnetic order parameter $L^2$ is related to the intensity of an antiferromagnetic Bragg peak (001 here), shown in the inset (from reference 12).*

Application below $T_N$ of the magnetic field on dielectric constant has been measured and compared to the magnetization curve exhibiting the steps and the plateaus associated to the magnetic transitions. We present in figure 2, the excess of dielectric constant $\varepsilon_H - \varepsilon_{sat}$ as function of magnetic field, where $\varepsilon_H$ is the value of dielectric constant at a given magnetic field H and $\varepsilon_{sat}$ is the value of dielectric constant at high magnetic field when magnetization gets saturated. The dielectric constant presents for a temperature of 10K two plateaus and two sharp peaks at H=0 and H=3.5T. This is very clear at 10K in which, by accident, the two plateaus are at the same level. It is still valid at 7K, but starts to be more complex at 4K, where some additional steps appear. We will focus in the following on the "high temperature" behaviour (7-25K).

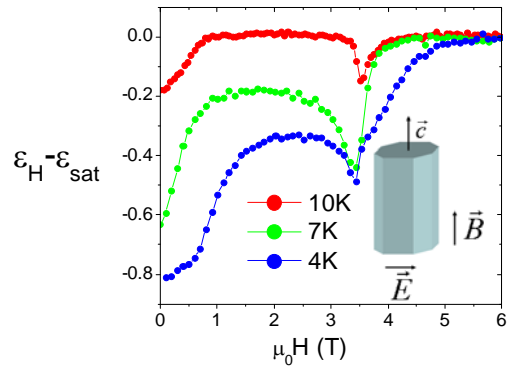

*Figure 2 : Magnetic field dependence of dielectric constant at different temperatures (4K, 7K and 10K).*

In order to analyse these data in the Landau analysis framework, we have collected the existing data on this sample. In reference [24], the magnetic field evolution of the magnetic peaks was studied by neutron diffraction on a single crystal. One can see on figure 4, extracted from this publication that *L* at 12K is first increasing, and then drops to zero at 3.6T. Correlatively, the magnetization *M* increasing up to $1/3M_s$ in a plateau and saturates to $M_s$ when *L* is dropping to zero. In addition, it was shown in this

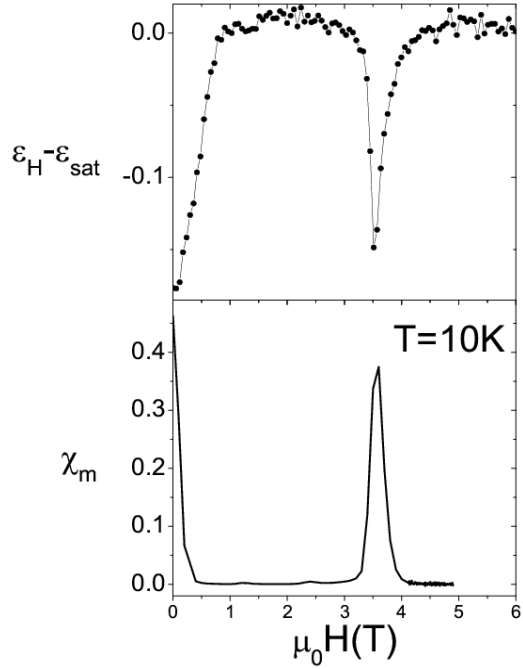

*Figure 3: Magnetic field dependence of magnetic susceptibility (lower part) and dielectric constant change (left upper part) at 10K. Magnetic and electric fields are applied along and perpendicular to the chains respectively.*

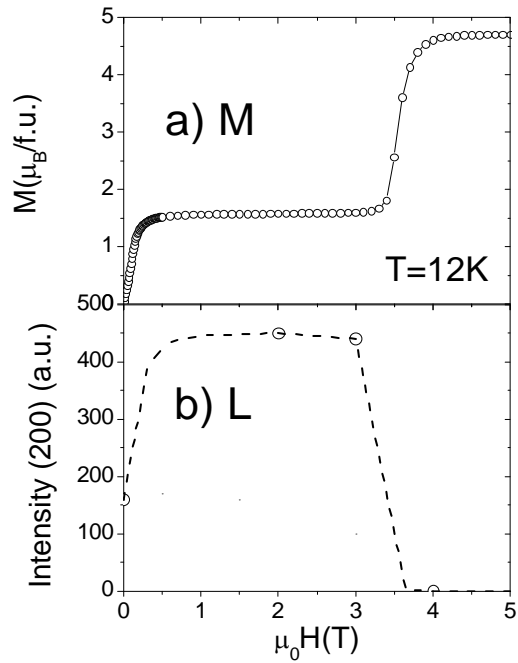

*Figure4: Field dependence of magnetic order parameters: a) magnetization at 10K and b) the neutron diffraction integrated intensity of the AFM peak (200) at 12K (Values extracted from reference24) ( dashed lines are guides to the eyes).*

work [24] that L depends strongly on temperature under magnetic field between 1 and 3.6 Tesla. Finally, the magnetic susceptibility $\chi_M = dM/dH$ was also measured at 10K (figure 3). The AF susceptibility $\chi_L = dL/dH$ was not measured in this publication (the number of points in H is very limited).

**IV Derivation of the free energy in presence of magnetic field**

The derivation of the free energy F in presence of magnetic field should be strongly modified from eq. 1 since the system is not purely antiferromagnetic. The following free energy has been used:

$$F = F_{AFM}(L) + F_{FM}(M) + \alpha P^2 - EP + \gamma P^2 L^2 + \lambda P^2 M^2 - MH + dH^2 L^2 \quad (4)$$

where the free energy of an antiferromagnet $F_{AFM}$ and that of a ferromagnet $F_{FM}$ are now both included since both kind of order parameters exist in this system under magnetic field. We have also added three terms $\lambda P^2 M^2 - MH + dH^2 L^2$ which represent the coupling with the applied field H and between P and M. With a similar derivation,

$$\frac{1}{\chi_e} = \frac{\partial^2 F}{\partial^2 P} = \alpha + \gamma L^2 + \lambda M^2 \quad (5)$$

Consequently, the dielectric constant can be developed for small $\gamma L^2 + \lambda M^2$ values:

$$\varepsilon = \chi_e + 1 = \frac{1}{\alpha + \gamma L^2 + \lambda M^2} + 1 \approx 1 + \frac{1}{\alpha}\left(1 - \frac{\gamma L^2}{\alpha} - \frac{\lambda M^2}{\alpha}\right) \quad (6)$$

But this equation does not explain the observed results since there is no peak in L or in M. In order to interpret the presence of these peaks, it is necessary to introduce a treatment of the fluctuations. A classical way to treat fluctuations in the frame of Landau theory is to use fluctuation-dissipation theorem which relates $\chi_M$ to $<M^2>-<M>^2$. ($<x>$ represents the averaging over the configuration space):

$$kT \chi_M = <M^2> - <M>^2 \quad (7)$$

By replacing $M^2$ by $<M^2>$, and then by its value from eq 7, (same for L),

$$\varepsilon = 1 + \frac{1}{\alpha} - \frac{\gamma \langle L \rangle^2}{\alpha^2} - \frac{\gamma kT\chi_L}{\alpha^2} - \frac{\lambda \langle M \rangle^2}{\alpha^2} - \frac{\lambda kT\chi_M}{\alpha^2} \quad (8)$$

Under application of magnetic field, the order parameter L is first increasing to reach a constant value in the first plateau, while *M* is increasing. At 3.6T, *L* drops to zero and *M* jumps to its saturated value. By accident, the variation of the two terms $\dfrac{\gamma \langle L \rangle^2}{\alpha^2} + \dfrac{\lambda \langle M \rangle^2}{\alpha^2}$ should cancel each other at 10K to explain why the two plateaus are at the same level.

In addition, there is an important contribution of the terms $\dfrac{\gamma kT\chi_L}{\alpha^2} + \dfrac{\lambda kT\chi_M}{\alpha^2}$ which are at the origin of the peaks in ε. $\chi_M$ is well known, but this is not the case of $\chi_L$. It is reasonable from the variation of L(H) to assume that it will also presents a peak at the same critical values of magnetic field.

Such an expression of the dielectric constant (eq 8) reproduces the observed behaviour, including the quantitative position of the peaks and the relative heights of the plateaus.

**V Conclusion**

In conclusion, a derivation of the free energy developed in the frame of the Landau theory (including two order parameters and a fluctuative part) shows that for such a complex magnetic state with different order parameters, the magnetic field dependence of the dielectric constant can be interpreted. There are many other aspects of the behaviour of the magnetization of this compound, related to its partially disordered state [15] which has not been studied here in detail. However, the spectacular observations (peaks in the dielectric constant related to peaks in magnetic susceptibility) open the routes to the application of the Landau analysis including fluctuations to other materials with magnetodielectric coupling.

Acknowledgements: The samples were kindly provided by Delphine Flahaut. We acknowledge many valuables discussions with Helmer Fjellvåg, Thomas Palstra Vincent Hardy and Raymond Frésard. Natalia Bellido acknowledges financial support to NOVELOX ESRT MEST-CT-2004-514237 and CNRS. We acknowledge financial support to the MaCoMuFi project.